\documentclass[manuscript]{acmart}
\AtBeginDocument{%
  \providecommand\BibTeX{{%
    \normalfont B\kern-0.5em{\scshape i\kern-0.25em b}\kern-0.8em\TeX}}}

\acmYear{2024}

\acmConference[With or Without Permission: Site-Specific Augmented Reality for Social Justice Workshop '24]{Workshop at CHI 2024}{May 11--16,
  2024}{Honolulu, HI}
\begin{document}

\title{The Clarkston AR Gateways Project: Anchoring Refugee Presence and Narratives in a Small Town}

\author{Joshua A. Fisher}
\email{joshua.fisher@bsu.edu}
\orcid{0000-0001-5628-5138}
\affiliation{%
  \institution{Ball State University}
  \city{Muncie, Indiana}
  \country{USA}
}

\author{Fernando Rochaix}
\email{frochaix@gsu.edu}
\orcid{0000-0002-7715-8514}
\affiliation{%
  \institution{Georgia State University, Perimeter College}
  \city{Decatur, Georgia}
  \country{USA}
  }

\renewcommand{\shortauthors}{Fisher and Rochaix}

\begin{abstract}
This paper outlines the Clarkston AR Gateways Project, a speculative process and artifact entering its second phase, where Augmented Reality (AR) will be used to amplify the diverse narratives of Clarkston, Georgia's refugee community. Focused on anchoring their stories and presence into the town's physical and digital landscapes, the project employs a participatory co-design approach, engaging directly with community members. This placemaking effort aims to uplift refugees by teaching them AR development skills that help them more autonomously express and elevate their voices through public art.  The result is hoped to be AR experiences that not only challenge prevailing narratives but also celebrate the tapestry of cultures in the small town. This work is supported through AR's unique affordance for users to situate  their experiences as interactive narratives within public spaces. Such site-specific AR interactive stories can encourage interactions within those spaces that shift how they are conceived, perceived, and experienced. This process of refugee-driven AR creation reflexively alters the space and affirms their presence and agency. The project's second phase aims to establish a model adaptable to diverse, refugee-inclusive communities, demonstrating how AR storytelling can be a powerful tool for cultural orientation and celebration.\end{abstract}



\begin{teaserfigure}
  \includegraphics[width=\textwidth]{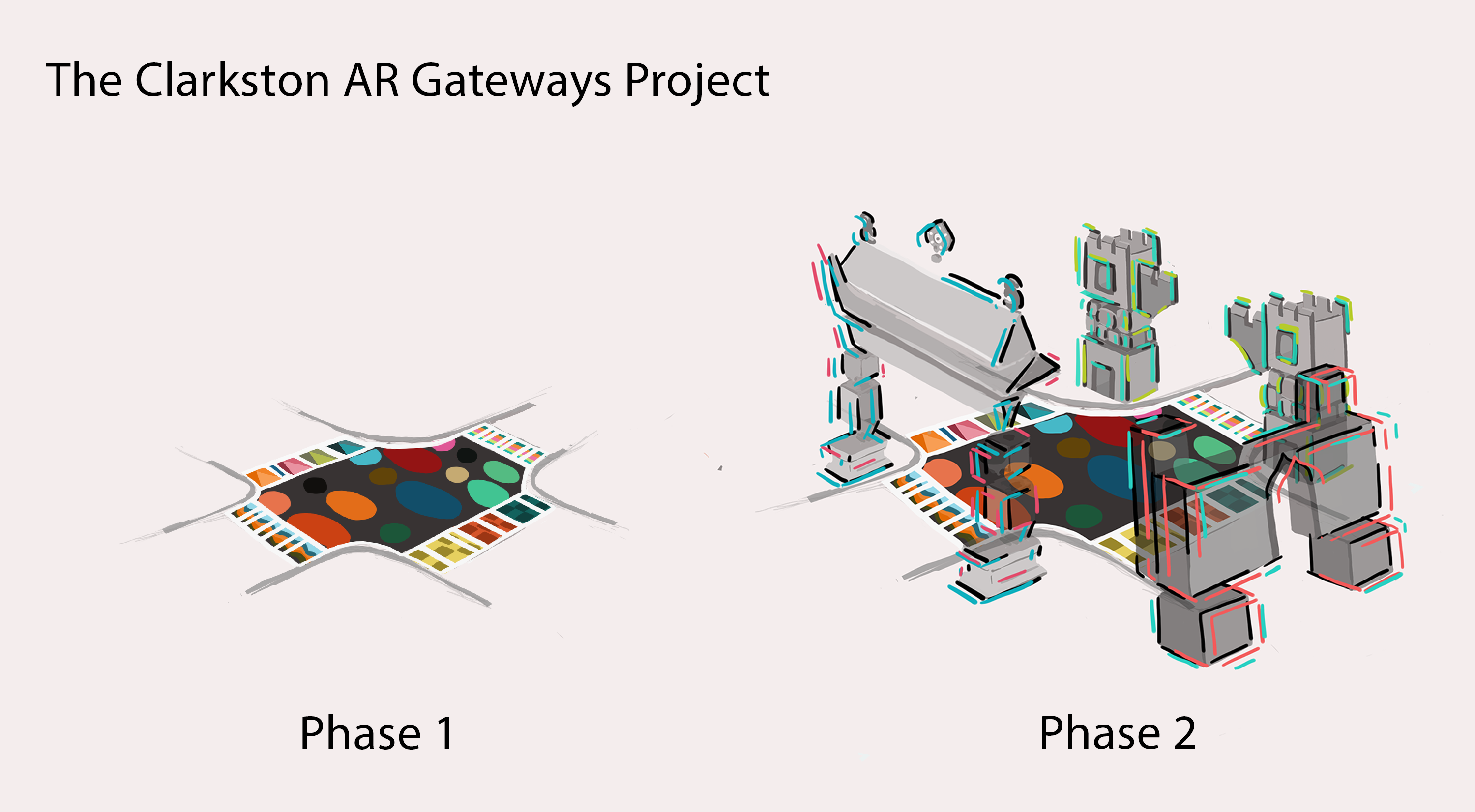}
  \caption{A sketch of the current state of the Clarkston AR Gateways project and where the project is going. Currently, phase 1 is complete. The paper discusses the development of phase 2, the addition of the AR.}
  \Description{A prismatic augmented reality arch in a city street. People are walking toward the arch.}
  \label{fig:teaser}
\end{teaserfigure}

\received{03/01/2024}
\received[accepted]{03/22/2024}

\maketitle

\section{The Speculative Artifact: A Clarkston AR Gateway}
Clarkston, Georgia, America's most diverse community per square mile, is known for its  refugee population from  55-plus countries\cite{yang2020refugee}. Students from the Clarkston community have collaborated in two previous physical placemaking initiatives co-designed by the local university. The Clarkston AR "Gateways" project expands upon this work by empowering the refugee community to engage in the discussions of public art, realizing their sustained presence in the community. Our current concept starts with an architectural armature as a postulate for interactive storytelling, a community's blank canvas for critical and playful expression. The choice of a liminal artifact is intentional and metaphorically flexible, enabling the refugees to root their life experiences in the town of Clarkston while providing a portal to their past, their cultural heritage, and the presence of their present in the new community. 

More often than not, in the United States, there has been a long history of "plop art," \cite{cartiere2022failures} where the art has little thematic connection to the site and the artist is the sole author of the work. The project's answer to this disconnect is to root the national and ethnic diversity in the composition of the gateways. Each is intended to render the legacy of the residents of Clarkston; each gateway is a window into immigrant symbolism, narratives, and experiences.

Participatory Design (PD), while addressing aspects of equity and decision-making, reaches its full potential through co-design \cite{kozubaev2019spaces}. This process not only amplifies the cultural narratives of diverse communities but also fosters technical empowerment. Specifically, the workshops aim to enhance the AR design and development capacities within the refugee community. By mastering these skills, refugees gain agency in curating their narratives and asserting their presence. This empowerment enables them to actively shape discussions surrounding public art in Clarkston, supporting the authentic representation of their cultures in and through the gateways. Moreover, proficiency in AR offers an alternative platform for engagement, reminiscent of the Situationists' concept of détournement \cite{debord1956methods}. It allows refugees to circumvent traditional channels and directly influence how their stories are woven into the community's cultural tapestry. This approach not only confronts the under representation of minority and non-citizen groups in public art discourse but also redefines the boundaries of how these discussions are conducted.

\subsection{The First Phase of the Project}
In 2019, Clarkston's Arts Council, with an economic development grant from the Georgia Placemaking Institute, initiated an art intervention to strengthen community bonds and add value to the area. This endeavor led to the commissioning of a crosswalk painting in the town center, completed on April 25, 2023, and intended to be a centerpiece in Clarkston's business district \label{Fig. 2}. The project, facilitated by our head designer as a researcher, aimed to foster inclusion and amplify marginalized voices. However, the process was not without its difficulties. During the research and development phase, the team faced the significant challenge of encapsulating the city's cultural diversity in a single functional art piece. Despite the noble intentions behind the crosswalk project, it encountered obstacles related to community participation and stringent funding deadlines. These challenges hindered the project's overarching goals of democratizing the development of civic cultural representations and building a more inclusive community narrative in public art.

The first issue was the city government's limited outreach to residents. Only ten responses were collected from a city-distributed online survey in a month. This gap highlights the challenge of achieving deeper participation. Such efforts can be perceived as extraction and exploitation. The PD process cannot be a one-sided knowledge extraction to foster inclusive, iterative dialogues\cite{simonsen2012routledge}. Within the city's selection committee, conflicts in personal taste influenced the final design, which incorporated bright colors and geometric shapes. As a result, while positioned as community-driven, traditional power dynamics persisted, limiting the project from living up to its potential. 

Lastly, the implementation process also faced challenges. The original grant required completing the crosswalk by an external deadline, which constrained possibilities for iterative co-design. In turn, this limited participation of the refugee groups did not result in their empowerment \cite{Huey2022Another}. Inadequate supplies further obstructed co-design efforts. Ultimately,  the team completed the work with limited input from residents, representing a fraction of the community's diversity. The project's first phase taught that realizing the democratizing ideals of PD depends on transferring authority over design decisions, timeframe, and resources directly to community members themselves. This guides the project's second phase as AR is integrated into the PD process. 
\begin{figure}
    \includegraphics[width=\textwidth]{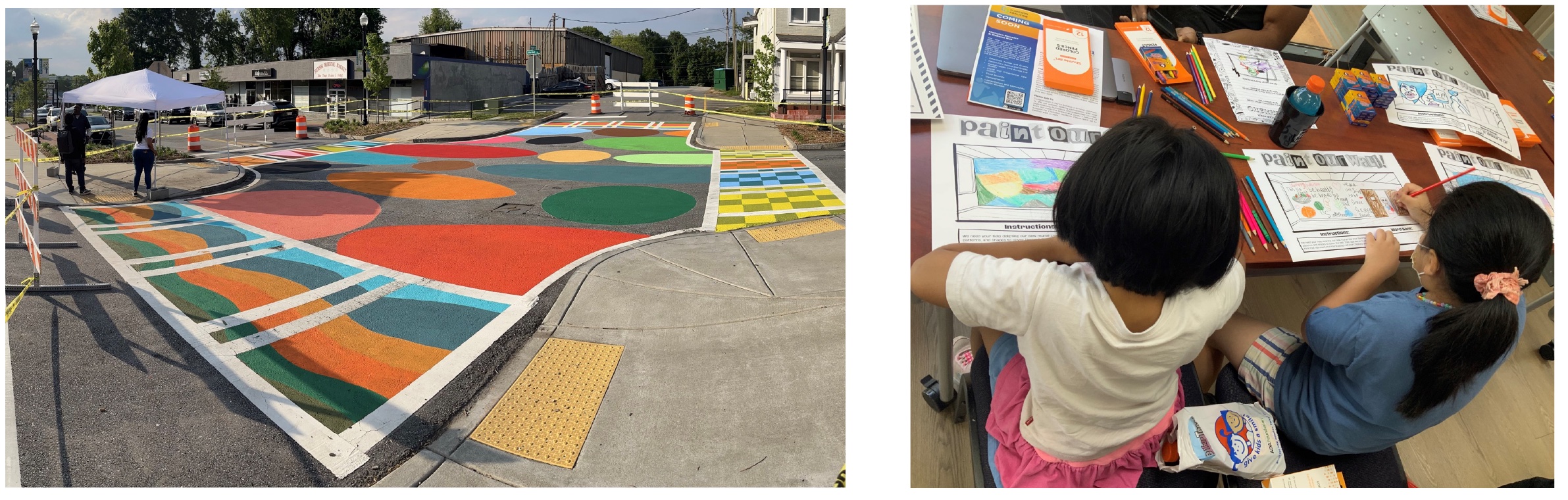}
Fig2:  \caption{Left: The Clarkston crosswalk mural was completed in April 2023. Right: First grade children from Indian Creek Elementary coloring in a preliminary design workshop. Their drawings were sources for our first iterative ideas of what would be appropriate for the space.}
\end{figure}
\section{Existing AR Work with Refugees}
Few AR experiences engage, interface, or seek to advocate for refugee communities. Working with refugees in Australia, researchers identified how AR's affordances might be used to support solutions and help individuals settle in their new environment \cite{Almohamed19}. Writing in 2019, they found that eXtended Reality (XR) did not and was not addressing the needs of the refugees who participated in their study. That said, the needs identified were one, to address issues of Post-Traumatic Stress Disorder; two, to find accurate information they need to settle in their new home; and three, cultural orientation in their new environments. The second phase of the  Clarkston AR Gateways project aligns with the last need through the guided co-design workshops that provide an opportunity for relationship and skills building during orientation. 

John Craig Freeman's Border Memorial, \textit{Frontera de Los Muertos,}\cite{Freeman2012} renders the deaths of undocumented immigrants, refugees fleeing violence in Central and South America, as skeletons that ascend into the sky at the locations across the United States' southern border where human remains were recovered. The work memorializes these deaths as dark tourism as being on site is a necessity \cite{fisher2018ethical}. The focus is on violence and death. The work does not address the presence, identities, or homes of those who passed as they attempted to cross the border. The result is a somber counter-monument \cite{auchter2020augmented} that speaks against the draconian immigration policies of the United States. In contrast to this critical and necessary confrontation, the Gateways Project operates from an ethics of care to guide a co-design process focused on integration, celebration, and healing. 

Another AR project, \textit{Journey with Me}, hosted by the University of British Columbia's Emerging Media Lab, followed the flight of Syrian refugees' journeys to settle in Vancouver. Five biographical stories are told through AR-enabled audio and movies.  The objective of the experience was to put students into the position of a refugee, "Joining content and form, students will be forced to make choices (or given a lack of choice), wait, and run/walk with the app to be able to progress the narrative of the experience. By the end of the experience, students should have a better understanding of both the physical and emotional journeys of Syrian refugees."  \cite{ubcJourneyWith} While laudatory, the belief that "By gamifying this experience, users can participate in understanding the emotional journey Syrian refugees have endured." is problematic. 

In 2018 and 2019, UK-based studio Collusion utilized AR and project mapping to tell the story of a Syrian refugee settling in King Lynn, in Norfolk, UK\cite{Ouda2022}.  Participants could stand and engage with the large-scale projections and seek out tracking images  with the letter R on them to discover new parts of the refugee's narrative. Involving the community further, Collusion furnished 300 children with digital lightboxes that interacted at site-specific locations. The isolated and continuous AR placemaking \cite{sharma2017framework} effort integrated the refugee experience into that of the contemporary city. 

Kanachai “Kit” Bencharongkul created NFT AR portraits of refugees and anecdotes from their personal stories. The work "Love is Boundless" was created in collaboration with the United Nations Refugee Agency. Its intent was "to fund talented refugee scholars to pursue a university degree or similar higher education. Net proceeds from this exclusive collection will be donated to UNHCR for life-changing opportunities for youth refugees around the world." \cite{morpheusLoveBoundless} While not site-specific, the work highlights the sociotechnical relationship between these kinds of emerging media initiatives, capitalistic structures, and the communities they are meant to benefit. The portraits and stories of refugees become bound up in the same assemblages that may have contributed to their experience to begin with. Their stories and narratives, not publicly shared, become commoditized financial instruments, even as the people they represent become distanced from the artifact representing them. The work quickly loses its context as an advocacy tool.

A project developed by the University of Toronto offers a personalized and interactive experience, focusing on intercultural exploration and celebrating refugee roots. This initiative, designed in collaboration with refugees, revolves around storytelling with a strong cultural dimension \cite{Sabie2023}. The narratives crafted by the refugees are intricately woven into the AR experience by attaching them to everyday objects like physical tea cups. When viewed in AR, these physical objects transform to reflect various cultural interpretations. For instance, a Senegalese tea cup triggers the rendering of an AR vibrant scene representing Senegalese culture. Users can also experience a mobile VR version of the culturally representative table. Users are active participants, engaging in traditional dining rituals and deepening their understanding of different cultures. This approach has been well-received and holds particular promise for the children of migrants, offering them a novel way to connect with and celebrate their heritage. Such initiatives highlight the power of AR in fostering cultural understanding and appreciation in diverse communities. 

The Clarkston AR Gateways project aims to transcend the gamification of refugee traumas or the commodification of their cultural heritage for entertainment. Instead, it emphasizes the importance of representation and establishing a tangible presence in space. This project adopts an approach of intentional reflexivity, empowering the refugee community to shape their future within the augmented environment by adding their stories as interactive experiences to the gateways. 
\section{\textbf{Using AR to Root a Story in a Community Space}}
"A lot of headlines when speaking about countries with conflict often focus on terrorism or war, or they focus on fear. What we want to do is uncover the stories about the people living there," says Rob Garcia, creator of an AR experience for World Refugee Day \cite{mobilesyrupWorldVision}. The refugees in Clarkston have taken those stories with them into their new homes. Their stories, in the form of location-based AR narratives, foster a more inclusive and understanding community ethos \cite{manfredinidiscussion}. Unlike non-location-based and non-AR works, the stories of refugees in AR engage directly with a heterotopia of real-world locations: one, in the present and unknown; another location, far away but both in the present and the past, containing their lives before, during, and after the conflict they have fled. Under the guidance of workshop organizers, refugees learn to employ interactive and spatial affordances \cite{murray2017hamlet} in the composition of the site's physical, cultural, social, political, and material aspects to tell that personal story \cite{fisher2021augmented}. This intersection of AR storytelling with social consciousness represents an opening of community spaces as inclusive and representative of diverse narratives \cite{jafarinaimi2015mrx, parvin2018doing}. In this manner, AR interactive storytelling presents a transformative avenue for actualizing spatial justice\cite{soja2013seeking, silva2022understanding, samuels2021building}.

Janet Murray's concept of transformation \cite{murray2017hamlet, murray2018research}  highlights how dramatic agency within interactive narratives can encourage users to change their behavior or perspectives. Unlike traditional narrative forms in which audiences suspend their disbelief, in interactive storytelling, users enact the narrative and so actively create belief in the narrative worlds. As discussed by Roth and Koenitz, the interplay of eudaimonic appreciation and narrative sense-making fosters a deeper connection between the narrative and the user's lived experience, leading to introspection and situated knowledge. When such narratives occur through site-specific AR, the process reinforces knowledge about community spaces.

The transformative power of site-specific AR stories lies in intertwining these diegetic changes within the heterotopic narrative world with transformations in the users' cognition and behavior. As per Endel Tulving's concept of episodic memory, the spatial-temporal context of storytelling activates memories of personal experiences, further deepening the connection between the narrative and the user. This kind of storytelling happens through real-time interaction with the AR interactive experiences on location\cite{spierling2017extensible, haahr2015literary, dionisio2017fragments}. Users exercise their agency and enact new behaviors by enabling real-time interaction within the narrative experience. For example, a user interaction in \textit{PokémonGo} is a new behavior for that space when the user travels to a new physical location where no one has caught AR Pokemon before. The interface guides the interactor's behavior across lived reality and AR to engage in that physical space in a new way. This new behavior can alter how Soja would say the space is conceived, perceived, and experienced \cite{soja1998thirdspace}. In other words, the football pitch becomes a Pokémon gym for insiders and outsiders because even those without access to AR are impacted by the behavior of those with access.  In this manner, physical and lived geographies become augmented through AR narratives for the entirety of the community, not just those with AR access. The interactions necessary to exercise agency in those narratives result in the enactment of new behaviors that alter those geographies. Using AR to change spaces this way can be liberating and has been discussed since OccupyAR  \cite{skwarek2018augmented} and the AR Art manifesto at the Venice Biennale Intervention in 2011 \cite{thiel2014critical}.  For the refugees, their personal stories change how the space is experienced.  The layering of their multiple storylines allows users to enact different perspectives and historical layers within a singular place. This complexity is integral to the kaleidoscopic understanding of the space that can foster transformative experiences. For designers, leveraging these spatial-temporal dimensions of AR, stories can enable both refugees and non-refugees to engage in interactions that are accessible, equitable, and inclusive, aligning with Soja's tenets of spatial justice\cite{soja2013seeking}.

\section{\textbf{Looking Ahead: Spatial Justice AR Storytelling in Practice}}
As the workshop enters its second phase, the researchers will consider the tenets of spatial justice as they co-design with refugees to create a place where movement within a community space can trigger personal, episodic memories and foster an authentic engagement in the Clarkston community. Central to AR's efficacy in achieving spatial justice is participants' need to gain technical literacy in AR, which is necessary for community membership in AR environments \cite{cameron2021designing}. The refugees of Clarkston have a right to weave their experiences into the physicality of their new environment and the social, cultural, and political aspects of their community\cite{alevizou2020civic, manfredinidiscussion}. First, the gateways will be accessible, allowing a plurality of users to enact the AR on location. Second, the participation in and with the gateways will need to be enacted equitably and include different modalities of engagement available to a diverse range of users and on-site non-users. Third, this inclusive engagement encourages users and non-users to address issues, obstacles, and injustices within the Clarkston community through narrative-based behaviors and interactions. Designing with these three tenets addresses community issues and injustices,  promoting spatial justice through narrative enactment with public art.  

The researchers intend to cooperatively develop templates, design docs, and manuals with the community on using AR for cultural heritage storytelling and placemaking. Such work will be approached with cultural sensitivity and co-creative practices. Community toolkits are not new interventions, and the majority intend to make research a tangible benefit, not after but during the equitable exchange of ideas and identities with participants. Hopefully, this approach process will support the fidelity and relevance of refugees' stories. 

During the summer of 2024, the team will conduct participatory design workshops with various Clarkston stakeholder groups (civic, religious, educational, artistic, etc.). These workshops will introduce the idea of public gateways to see if this imagined artifact can indeed carry the needs and design solutions requested by various community constituents. We see this as a \textit{tabula rasa}; the end armature could be something other than a gate. This approach validates refugees' experiences and fosters a sense of ownership of space and place within Clarkston. Working directly with the refugees helps the Clarkston community utilize AR to resist a future where "new digital technologies have strengthened the hegemonic position of dominant actors by expanding the capability of their apparatuses to control and disempower." \cite{manfredini2021critical}. Instead, AR will be utilized to root their rich heritage and immigrant experience within the community. This digital-physical-socio-cultural nucleation empowers them to claim their spatial agency and presence, educating and engaging the broader community in a dialogue of understanding and compassion. 

\begin{acks}
We extend our heartfelt gratitude to the city and people of Clarkston, Georgia, for their openness and trust in collaborating with us to share their unique stories. We are equally thankful for the dedicated efforts of our student-led undergraduate design team, whose diligent work in gathering preliminary data was crucial for the development of our first speculative artifact. Lastly, special recognition is due to Milan Robinson for her exceptional contribution to our article through her skilled work on the image at the top of the article. 
\end{acks}

\bibliographystyle{ACM-Reference-Format}
\bibliography{sample-base}

\appendix

\end{document}